\documentstyle[12pt,cite,epsfig,psfrag]{article}

\def\be{\begin{equation}}
\def\ee{\end{equation}}
\def\bea{\begin{eqnarray}}
\def\eea{\end{eqnarray}}

\newcommand{\beqal}{\begin{eqnarray}\label}
\newcommand{\beqa}{\begin{eqnarray}}
\newcommand{\eeqa}{\end{eqnarray}}

\begin{document}

\begin{titlepage}
\begin{center}
\vskip .2in

{\Large \bf Viscosity to entropy ratio at extremality}
\vskip .5in

{\bf Sayan K. Chakrabarti\footnote{e-mail: sayan@iopb.res.in}, 
\bf Sachin Jain\footnote{e-mail: sachjain@iopb.res.in},
\bf Sudipta 
Mukherji\footnote{e-mail: mukherji@iopb.res.in}\\
\vskip .1in
{\em Institute of Physics,\\
Bhubaneswar 751~005, India.}}
\end{center}

\begin{center} {\bf ABSTRACT}

\end{center}
\begin{quotation}\noindent
\baselineskip 15pt
Assuming gauge theory realization at the boundary, we show that the
viscosity to entropy ratio is ${1/(4 \pi)}$ where the bulk is 
represented by a large class of extremal black 
holes in anti-deSitter space. In particular, this class includes
multiple R-charged black holes in various dimensions. 
\end{quotation}
\vskip 2in
October 2009\\
\end{titlepage}
\vfill
\eject

Extremal black holes are special in many ways. Often, various computations 
tend to break down as one tries to extract out results associated with extremal black 
holes via  `extremal limit' of non-extremal black holes. One such example recently has 
appeared in the calculation of shear viscosity ($\eta$) to entropy ($s$) ratio for gauge 
theory that is dual to extremal bulk geometry. In particular, in
the low frequency limit ( $\omega \rightarrow 0$ limit or in other words the 
IR limit of the boundary gauge theory), used for example in \cite{Son:2006em}, 
the perturbation in $\omega$ breaks down. 
In \cite{Faulkner:2009wj}, a prescription was given which can be used to 
treat these extremal holes.
Subsequently, in \cite{Edalati:2009bi}\footnote{For certain class of black holes on 
AdS$_5$, a discussion on $\eta/s$ can be found in \cite{Imeroni:2009cs}.}, following this 
prescription, $\eta/s$ 
was computed for four dimensional Reissner-N\"ordtstrom black holes in AdS. The result
turned out to be $1/(4 \pi)$; same as their non-extremal partners. It was further 
argued that, regardless of the dimensions of space-time, the result would remain 
unchanged for Reissner-N\"ordtstrom black hole.

Encouraged by these developments, in this note, we provide a 
computation of $\eta/s$
for a generic extremal black hole in arbitrary dimensions having
metric of the form 
\begin{equation}
ds_{d+1}^2 = \bar g_{tt} dt^2 + \bar g_{uu} du^2 + \bar g_{ij} dx^i dx^j,
\label{chmetric}
\end{equation}
with
\begin{equation}
\bar g_{tt} = - f(u) A_1(u), ~~\bar g_{uu} = A_2(u) f^{-1}(u),~~f(u)=(1-u)^2 V(u).
\end{equation}
In terms of coordinate 
$u$, the horizon is located at $u = 1$ while the boundary is at $u=0$. 
We take functions $A_1(u), A_2(u)$ and $V(u)$ to be finite on the horizon.
Extremal nature of this geometry shows up in the double pole of $f^{-1}(u)$ 
at the horizon. We assume that the metric asymptotes to AdS and, 
on the boundary, we have an associated gauge theory. Among others,
this class of metric includes generic R-charged black holes in five dimensions 
\cite{Behrndt:1998jd}. The boundary gauge theory is then expected to be strongly coupled 
SYM on $R^4$ in the presence of three non-zero chemical potentials. 
This metric also includes multiple R-charged 
black holes in four and seven dimensions \cite{Cvetic:1999xp}. However, the gauge 
theories at their boundaries are less clearly understood. Besides the metric 
(\ref{chmetric}), there are gauge fields and scalars. The detail forms of these 
quantities will not be required for our present work. The geometry is characterised 
by the fact that its entropy is finite even though the temperature  is zero 
\cite{Lu:2009gj, Jain:2009uj}. We now proceed to compute $\eta/ s$ 
associated with this geometry.

First, to compute the shear viscosity, one considers some specific fluctuations of the 
metric 
and uses Kubo formula. This formula relates the viscosity to the correlation function
of the stress-energy tensor at zero spatial momentum. We will avoid the details here. 
They can be found in \cite{Son:2002sd,Son:2006em,Edalati:2009bi}. Writing  
$g_{\mu\nu} = \bar g_{\mu\nu} + h_{\mu\nu}$ with $\bar g_{\mu\nu}$ given in 
(\ref{chmetric}), and Einstein equation leads to the
following equation for ${h^{x}}_{y}$ (which turns out to be same as 
that of  massless real scalar field. In what follows, we call it  $\Phi$.)
\begin{equation}
\partial_\mu\Big({\sqrt{-\bar g}} ~\bar g^{\mu\nu} \partial_{\nu}\Big)\Phi = 0.
\end{equation}
 Further, using the 
ansatz $\Phi = e^{-i \omega t} \phi(u)$, we get
\begin{equation}
\partial_u^2 \phi + \partial_u {\rm ln}(\bar g^{uu} {\sqrt{-\bar g}}) \partial_u\phi 
- \frac{\bar g^{tt}}{\bar g^{uu}} \omega^2 \phi = 0.
\end{equation}
Using explicit forms of $ \bar g^{tt}, \bar g^{uu}$, we finally reach at an equation of 
the form
\begin{equation}
\partial_u^2 \phi + \partial_u {\rm ln}\Big(\frac{ {\sqrt{-\bar g}}f}{A_2}\Big) 
\partial_u\phi
- \frac{\omega^2 A_2}{f^2 A_1} \phi = 0.
\label{phieq}
\end{equation}
 We solve the above equation in the inner region (near the horizon) as well as in the outer
region (away from the horizon). We then match both at the  so called matching 
region \cite{Edalati:2009bi}. We first look for solution in the inner region.

Due to the double pole singularity in $f^{-1}(u)$, the usual low frequency ($\omega$)
expansion of $\phi$ becomes subtle \cite{Faulkner:2009wj,Edalati:2009bi}. One then 
defines $\xi$ as $u = 1 - \omega/\xi $ and organises the solution as an expansion in 
$\omega$ where $\omega\rightarrow 0$ and $\xi\rightarrow 0$ in such a way that $\omega/\xi\rightarrow 0$, see 
\cite{Edalati:2009bi} for details. The equation (\ref{phieq}) then simplifies to 
 (keeping only zeroth order in $\omega$)
\begin{equation}
 \frac{\partial^2 \phi}{\partial \xi^2} + \frac{A}{V^2}\phi =0,
\end{equation}
where, 
\begin{equation}
A = \frac{A_2(u)}{A_1(u)}\Big |_{u =1},\,\,\,\,V=V(u)|_{u\to 1}.
\end{equation}
Next, defining $\alpha = \frac{\sqrt{A}}{V} \xi$ above 
equation reduces to standard form in $AdS_2$
\begin{equation}
\frac{\partial^2 \phi}{\partial \alpha^2} + \phi =0.
\end{equation}
The incoming wave solution is then
\begin{equation}
\phi_{in} = a_{I}^0 e^{i \alpha} \sim a_I^0 ( 1 + i \alpha)
= a_I^0 + \frac{g({\omega})}{1 - u} a_I^0.
\label{innerreg}
\end{equation}
with 
\begin{equation}
g(\omega) = \frac{i {\sqrt{A}}}{V} \omega.
\label{gomega}
\end{equation}
In (\ref{innerreg}), $a_I^0$ is a constant. 
Since (\ref{innerreg}) represents incoming solution near the horizon, this
is often called the solution in the inner region or the solution in the
IR of the boundary gauge theory.

As for the solution in the outer region or in other words away from the 
horizon, we go back to equation (\ref{phieq}). Note that in this case, to 
zeroth order in $\omega $, 
we get 
\begin{equation}
\partial_u^2 \phi + \partial_u {\rm ln}(\bar g^{uu} {\sqrt{-\bar g}}) \partial_u\phi =0. 
\end{equation}
Integrating once, 
\begin{equation}
\partial_u \phi = c_1 \frac{\bar g_{uu}}{\sqrt{-\bar g}}, 
\end{equation}
where $c_1$ is a constant.
This implies
\begin{equation}
\phi_{out} = c_2 + c_1 \Big[ \frac{F(u)}{(1 - u) V(u)} - \frac{F^\prime(u)}{V(u)} 
{\rm{ln}} (1 - u)
+ \int \Big(\frac{F^\prime(u)}{V(u)}\Big)^\prime {\rm{ln}} (1 - u) ~du\Big],
\label{outerreg1}
\end{equation}
where, 
\begin{equation}
 F(u) = \frac{A_2(u)}{\sqrt{-\bar g}}.
\end{equation}
In (\ref{outerreg1}), $c_2$ is an integration constant.
Note further that $F(u)$ is finite on the horizon. In order to get 
the complete low frequency profile of 
$\phi$, we need to match (\ref{outerreg1}) and (\ref{innerreg}) at $u \rightarrow 1$.
Outer region solution gives
\begin{equation}
\phi_{out} = c_2 + c_1 B + \frac{c_1  F}{V (1 - u)},
\label{uout}
\end{equation}
with $F = F(u)|_{u\rightarrow 1}$
and 
\begin{equation}
 B = \Big[  - \frac{F^\prime(u)}{V(u)} 
{\rm{ln}} (1 - u)
+ \int \Big(\frac{F^\prime(u)}{V(u)}\Big)^\prime {\rm{ln}} (1 - u) ~du\Big]_{u\rightarrow 1}
\end{equation}
Now comparing (\ref{uout}) with (\ref{innerreg}), we get 
\begin{equation}
c_1 = \frac{V g(\omega)}{F} a_I^0, ~~ c_2 = a_I^0 \Big( 1 - \frac{B V g(\omega)}{F}\Big).
\end{equation}
Substituting these constants in (\ref{outerreg1}) 
\begin{eqnarray}
\phi_{out} &=& a_I^0 \Big( 1 - \frac{B V g(\omega)}{F}\Big) + 
\frac{V g(\omega)}{F} a_I^0 \Big[ \frac{F(u)}{(1 - u) V(u)} - \frac{F^\prime(u)}{V(u)} 
{\rm{ln}} (1 - u) \nonumber \\
&+& \int \Big(\frac{F^\prime(u)}{V(u)}\Big)^\prime {\rm{ln}} (1 - u) ~du\Big].
\end{eqnarray}
Hence
\begin{equation}
\partial_u \phi_{out} = \frac{V g(\omega)}{F} a_I^0 \frac{\bar g_{uu}}{\sqrt{-\bar g}}. 
\label{outd}
\end{equation}
Now it is straightforward to compute the boundary action and then the 
correlation function following \cite{Son:2002sd}. As for the boundary action, we get
\begin{equation}
S_{\rm boundary} = -\frac{1}{2}\frac{1}{16 \pi G}\Big[ \bar g^{uu}\sqrt{-\bar g}\phi_{out} \partial_u \phi_{out} \Big]_{u=\epsilon\rightarrow 0}
=- \frac{g(\omega) V (a_I^0)^2}{32 \pi G F}.
\end{equation}
Hence, to first order in $\omega$
\begin{equation}
G_{xy,xy} = \frac{\partial S_{boundary}}{\partial a_I^0 \partial a_I^0}
= -\frac{g(\omega) V}{16 \pi G F}
=-\frac{i {\sqrt{A}} \omega}{16 \pi G F}.
\end{equation}
Here $G$ is the $d+1$ dimensional Newton's constant.
In the last line we have used the form of $g(\omega)$ given in 
(\ref{gomega}). Now the Kubo formula gives us the shear viscosity \footnote{
We observe that the form of $\eta$ is same as that obtained in the non-extremal
cases \cite{Iqbal:2008by, McGreevy:2009xe}. The structural similarity leads us to 
speculate that there 
might be a membrane paradigm like prescription for extremal black holes having 
non-zero entropy.} 
\begin{equation}
\eta = \frac{\sqrt{A}}{16 \pi G F} =
\frac{1}{16 \pi G}\left[\sqrt{ 
\frac{{-\bar g}}{{\bar g_{uu} \bar g_{tt}}}}\right]_{u=1}.
\label{eta23}
\end{equation}

Since the entropy density of black hole is given by 
\begin{equation}
s = \frac{\sqrt{{\rm det}\bar g_{ij}}|_{u=1}}{4 G}=
\frac{1}{4 G}\Big[\sqrt{ \frac{{-\bar g}}{{\bar g_{uu} \bar g_{tt}}}}\Big]_{u=1}
\end{equation}
we get 
\begin{equation}
\frac{\eta}{s} = \frac{1}{4 \pi}.
\end{equation}
As an illustrative example, we now consider R-charged black hole in seven dimensions and 
try to see how it fits into the above scheme. This black hole can carry at most two 
independent R-charges. The metric, in the extremal limit, is given by 
\cite{Cvetic:1999xp}
\begin{equation}
ds_7^2 =  \frac{4 H^{1/5} (\pi l T_0)^2 }{9 u} \Big( -\frac{f(u)}{H(u)} dt^2 + 
dx_1^2+\cdots+dx_5^2 \Big) + \frac{l^2 H(u)^{1/5}}{4 f(u) u^2} du^2,
\label{twoch}
\end{equation}
where, 
\begin{eqnarray}
&&f(u)=  (1+2u + k_1 k_2 u^2) (1 - u)^2,\nonumber\\
&& H(u) = ( 1 + k_1 u^2) (1 + k_2 u^2).
\end{eqnarray}
In writing down the above geometry, we have used extremality condition 
\begin{equation}
3+k_1+k_2= k_1 k_2.
\end{equation} 
The metric has a double pole at $u = 1$. 
Comparing with our general notations used earlier, we find that for (\ref{twoch}), 
\begin{eqnarray}
&& \sqrt {-\bar g}=\frac{l H^{1/5} (\frac{2 \pi l T_0)}{3})^6 }{2 u^4}, ~~A_1(u) = 
\frac{4 (\pi l T_0)^2 H^{-4/5}}{9 u}, ~~A_2 (u)  = \frac{H^{1/5} l^2}{4 
u^2},\nonumber\\
&& V(u) = 1+2u + k_1 k_2 u^2,
~~F(u) = \frac{l u^2}{2 (\frac{2 \pi l T_0}{3})^6}.
\end{eqnarray}
Note that all these quantities are finite at the horizon. From our previous 
discussion, it then immediately follows that $\eta/s = 1/(4 \pi)$.

Encouraged by the membrane paradigm arguments (for details see for 
example \cite{Iqbal:2008by, 
McGreevy:2009xe}), we end this note with an alternative derivation of $\eta$ 
for extremal black hole. Consider the bulk action for a massless scalar $\Phi$:
\begin{eqnarray}
S_{\rm bulk}=\frac{1}{2}\int d^{d+1}x
\sqrt{-\bar g}\frac{\partial_A\Phi\partial^A\Phi}{16\pi G}
\end{eqnarray}
Using linear response theory one can write the transport coefficient as 
\bea
\chi =\lim_{\omega\rightarrow 0}\lim_{u\rightarrow 0}
\left(\frac{\Pi_{\Phi}(u,\omega)}{i\omega\Phi(u, \omega)}\right),
\eea
where $\Pi_{\Phi}(u,t)=\frac{\partial \mathcal{L}_{\rm{bulk}}}
{\partial(\partial_u\Phi)}$\cite{McGreevy:2009xe}. Note that  $\Pi_{\Phi}(u,\omega)$ is the 
Fourier transform of the function $\Pi_{\Phi}(u,t)$. If we take $\Phi(u,t) = 
h^x_y$, then we get $\eta$ as the transport coefficient. Following our previous 
discussion, we note that the field momentum is of the form
\bea
\Pi_{\Phi}(u,\omega)=\frac{\sqrt{-\bar g}}{16\pi G}\bar g^{uu}\partial_u\phi.
\eea
Now using the fact that
\bea
\partial_u\phi_I=i\frac{\partial \alpha}{\partial
u}\phi_I=i\frac{\sqrt{A}}{V}\frac{\omega}{(1-u)^2}\phi_I,\label{eqn32}
\eea
and (\ref{outd}), we see
\bea
\eta = \lim_{\omega\rightarrow 0}\lim_{u\rightarrow 0}
\left(\frac{\Pi_{\Phi}(u,\omega)}{i\omega\Phi(u, \omega)}\right)
= \lim_{\omega\rightarrow 0}\lim_{u\rightarrow 1}
\left(\frac{\Pi_{\Phi}(u,\omega)}{i\omega\Phi(u, \omega)}\right)
= \frac{1}{16\pi G}\sqrt{\frac{-\bar g}{\bar g_{uu}\bar g_{tt}}}\Big |_{u\rightarrow 1}.
\eea
This is same as what we got previously (\ref{eta23}). To evaluate the above expression, 
we have used (\ref{eqn32}) for  $\phi$  in $u\rightarrow 1$ region
and (\ref{outd}) for $u\rightarrow 0$ region. 

To conclude, we have shown that the viscosity to entropy ratio is insensitive to many
details of the extremal AdS black hole geometry. For our computation, we only required
the double pole nature of $\bar g_{uu}$ and double zero of $\bar g_{tt}$ at the horizon. Rest of the quantities
associated with the metric are only assumed to be finite and non-zero on the horizon.
Given these information, we argued that analytic expression for shear viscosity and the
viscosity to entropy ratio remain same as that of many non-extremal black holes where near
horizon geometry is radically different. We have also observed that a membrane paradigm
like arguments for computation of the shear viscosity go through in the extremal case
with double pole nature of metric, even though the computations of [10] seem to depend
crucially on the the single pole nature of the geometry. This leads us to believe that, in
spite of differences in the horizon structure between non-extremal and extremal black hole
geometries, the field momentum relevant for determination of the shear viscosity, is blind
to such differences.

\vspace{2cm}
\noindent
{\bf Note:}  A day before posting our paper on the arXiv, references  
\cite{Paulos:2009yk} and \cite{Cai:2009zn} appeared which has some overlap with the present paper.

\end{document}